\documentclass[11pt]{article}
\usepackage{amsmath, amsthm, amsfonts}
\usepackage[margin=1in,footskip=0.25in]{geometry}
\usepackage{makecell}
\usepackage{multicol}
\usepackage{graphicx}
\theoremstyle{theorem}

\usepackage{hyperref}
\usepackage{cite}

\newtheoremstyle{defi}
  {10pt}          
  {10pt}  
  {\rm}  
  {\parindent}     
  {\bf}  
  {. }    
  { }    
  {}     
\theoremstyle{defi}

\begin{document}

\date{}

\title{\bf Dark energy and its manifestations}
\author{Koijam Manihar Singh$^{1}$, Gauranga C. Samanta$^{2}$ \\
 $^{1}$ ICFAI University Tripura, Kamalghat, Mohanpur-799210, Tripura, India \\
 $^{2}$Department of Mathematics,
BITS Pilani K K Birla Goa Campus,
Goa-403726, India,\\ drmanihar@rediffmail.com\\ gauranga81@gmail.com}

\maketitle


\begin{abstract}  In a four dimensional manifold formalism we study the evolutionary behavior as well as the ultimate fate of the universe, in the course
of which the contribution of dark energy in these phases are investigated. At one stage we get a situation (a condition) where the dark energy contained dominates other types of energies available in this universe. In the model universes we obtain here the dark energy is found to be of $\Lambda$CDM
and quintessence types-which bear testimony to being real universes. In one of the cases where the equation of state between the fluid pressure and density is of the type of the van der Waals equation, it is found that our universe may end in dust. And, also, it is seen that the behavior of the deceleration parameter is almost compatible with the recent observation.
\end{abstract}



\section{Introduction}
Recently, various observations suggested that our universe is experiencing an accelerated expansion phase \cite{1}. We believe that this accelerated expansion of the universe may be the result of some mysterious dark energy. Many research papers have been published yet and rigorous research is going on, to explain the origin of this mysterious dark energy. Nevertheless, the source of this dark energy is an open problem for the present time. The cosmological constant $\Lambda$ is the simplest candidate to explain dark energy, and it is characterized by the parameter $\omega=-1$ of the equation of state $p=\omega \rho$ and constant energy density $\rho$ \cite{2}. There are two fundamental physical parameters in Einstein's field equations, namely, the gravitational coupling $G$ and the cosmological constant $\Lambda$, which are usually assumed to be constants \cite{3}. In the Einstein field equations the Newtonian gravitational constant $G$ acts as a coupling constant between matter and geometry; and this $G$ can behave also as a function of cosmic time as proposed by \cite{4}. Subsequently, in the past few decades, several suggestions have been proposed for $G$ to vary with time based on different arguments; and many other extensions of Einstein's theory with variable $G$ have also been proposed in order to achieve a possible unification of gravitation and other forces of nature\cite{5, 6, 7}.

In early stage, the cosmological constant was appealed twice in modern cosmology. Einstein was the first researcher who introduced cosmological constant to construct a static model of universe \cite{Bernstein}. The universe possesses a non-zero cosmological constant, which is suggested by the observational data \cite{8}. In the last few years the cosmological constant problems have been studied seriously, which promoted researchers to look in different ways in the names of dynamical dark energy models \cite{Amendola, Kiefer, Sahni1, Singh1, Singh2, Singh3, Singh4, Singh5, Priyokumar, Singh6, Supriya} \bibliographystyle{IEEEtran}, modified gravitational theories \cite{Nojiri, Nojiri1, Nojiri2,Felice, Sotiriou, Capozziello, Cai, Nojiri3, Bamba, Bamba1, Bamba2, Samanta, Samanta1, Samanta2, Samanta3, Samanta4, Oikonomou}. Consequently, the time varying $\Lambda$ theory has attracted the attention of researchers in recent years to explore new findings. Commonly, the time dependent cosmological models can be developed to investigate the expansion history of the universe. The coincidence problem may be explained by the class of cosmological models, which is coupled with time dependent cosmological constant $\Lambda=\Lambda(t)$.

 Therefore, the cosmological model with time dependent cosmological constant might be reflected to be an interesting problem to investigate the dynamics of dark energy as well as to focus on the expansion history of the universe.

\section{Various equations and exact solutions}
Let $M$ be a $4$-dimensional manifold equipped with a metric $\bar{g}_{\mu\nu}$, which determines
the space-time interval as follows:
\begin{equation}\label{1}
  d\bar{s}^2=\bar{g}_{\mu\nu}dx^{\mu}dx^{\nu}=dt^2-a^2(t)\bigg[\frac{dr^2}{1-kr^2}+r^2(d\theta^2+\sin^2\theta d\phi^2) \bigg]
\end{equation}
where $k=-1, 0, 1$ is the curvature.

Now the corresponding Einstein field equations with time varying $\Lambda$ and $G$ give
\begin{equation}\label{2}
  8\pi G\rho+\Lambda=3\frac{\dot{a}^2}{a^2}+3\frac{k}{a^2}
\end{equation}
and
\begin{equation}\label{3}
  8\pi G p-\Lambda=-2\frac{\ddot{a}}{a}-\frac{\dot{a}^2}{a^2}-\frac{k}{a^2}.
\end{equation}
Here the continuity equation takes the form
\begin{equation}\label{4}
  \dot{\rho}+3H(p+\rho)=0
\end{equation}
where $H$ is the Hubble parameter. \\ \\
\textbf{Case-I:}

We take the equation of state to be
\begin{equation}\label{5}
  \frac{p}{\rho}=-\omega.
\end{equation}
Now adding up \eqref{2} and \eqref{3} we have
\begin{equation}\label{6}
  8\pi G (p+\rho)=2\frac{\dot{a}^2}{a^2}-2\frac{\ddot{a}}{a}+2\frac{k}{a^2}
\end{equation}
Also \eqref{4} and \eqref{6} give
\begin{equation}\label{7}
  -8\pi G\frac{a}{\dot{a}}\dot{\rho}=6\frac{\dot{a}^2}{a^2}-6\frac{\ddot{a}}{a}+6\frac{k}{a^2}
\end{equation}
Again differentiations both sides of \eqref{2} we get
\begin{equation}\label{8}
  8\pi G \dot{\rho}+8\pi \dot{G}\rho+\dot{\Lambda}=6\frac{\dot{a}\ddot{a}}{a^2}-6\frac{\dot{a}^3}{a^3}-6k\frac{\dot{a}}{a^3}
\end{equation}
And from \eqref{6} we have
\begin{equation}\label{9}
  -3\frac{\dot{a}}{a}[8\pi G(p+\rho)]=6\frac{\dot{a}\ddot{a}}{a^2}-6\frac{\dot{a}^3}{a^3}-6k\frac{\dot{a}}{a^3}
\end{equation}
From \eqref{8} and \eqref{9},
\begin{equation}\label{10}
  8\pi G \dot{\rho}+8\pi\dot{G}\rho+\dot{\Lambda}=-3\frac{\dot{a}}{a}[8\pi G(p+\rho)]
\end{equation}
Using \eqref{4} and \eqref{10} we have
\begin{equation}\label{11}
  8\pi \dot{G}\rho+\dot{\Lambda}=0
\end{equation}
Again \eqref{4} and \eqref{5} give
\begin{equation}\label{12}
  \rho=a_0a^{3(\omega-1)},
\end{equation}
where $a_0$ is an arbitrary constant.
Now from \eqref{5} and \eqref{12}, we have
\begin{equation}\label{13}
  p=-\omega a_0a^{3(\omega-1)}
\end{equation}
Then from \eqref{4}, \eqref{5} and \eqref{6} we get
\begin{equation}\label{14}
  -8\pi G(\omega-1)a_0 a^{3(\omega-1)}=2\frac{\dot{a}^2}{a^2}-2\frac{\ddot{a}}{a}+2\frac{k}{a^2}
\end{equation}
Without loss of generality we separate \eqref{14} into two equations
\begin{equation}\label{15}
  3\frac{\dot{a}^2}{a^2}-3\frac{\ddot{a}}{a}=-c
\end{equation}
and
\begin{equation}\label{16}
  -8\pi G(\omega-1)a_0 a^{4(\omega-1)}=-c+ 3\frac{k}{a^2}
\end{equation}
where $c$ is a separation constant. \\
Now \eqref{15} gives,
\begin{equation}\label{17}
  a=e^{\frac{ct^2}{2}+c_0t+c_1},
\end{equation}
where $c_0$ and $c_1$ are arbitrary constants. Here the scale factor expands rapidly with time and no singularity is observed within finite time. And from \eqref{16} we have
\begin{equation}\label{18}
  G=\frac{9}{8\pi a_0(\omega-1)}[c-ke^{-ct^2-2c_0t-2c_1}]e^{-3(\omega-1)(\frac{ct^2}{2}+c_0t+c_1)}
\end{equation}
The gravitational constant $G$ becomes infinity for $\omega=1$ and it converges to $zero$ as $t\to\infty$ for $c=0$, however it diverse to infinity as $t\to\infty$ for non zero $c$. Hence we conclude that, for $c=0$, there is no importance of gravitational constant $G$ in an infinite time.

Now using \eqref{2}
\begin{equation}\label{19}
  \Lambda=\frac{2\omega+1}{\omega-1}ke^{-ct^2-2c_0t-2c_1}+2(ct+c_0)^2-\frac{3c}{\omega-1}
\end{equation}
Here, the deceleration parameter
\begin{equation}\label{20}
  q=-1-\frac{c}{(ct+c_0)^2}
\end{equation}
\begin{figure}
  \centering
  \includegraphics[width=10cm,height=8cm]{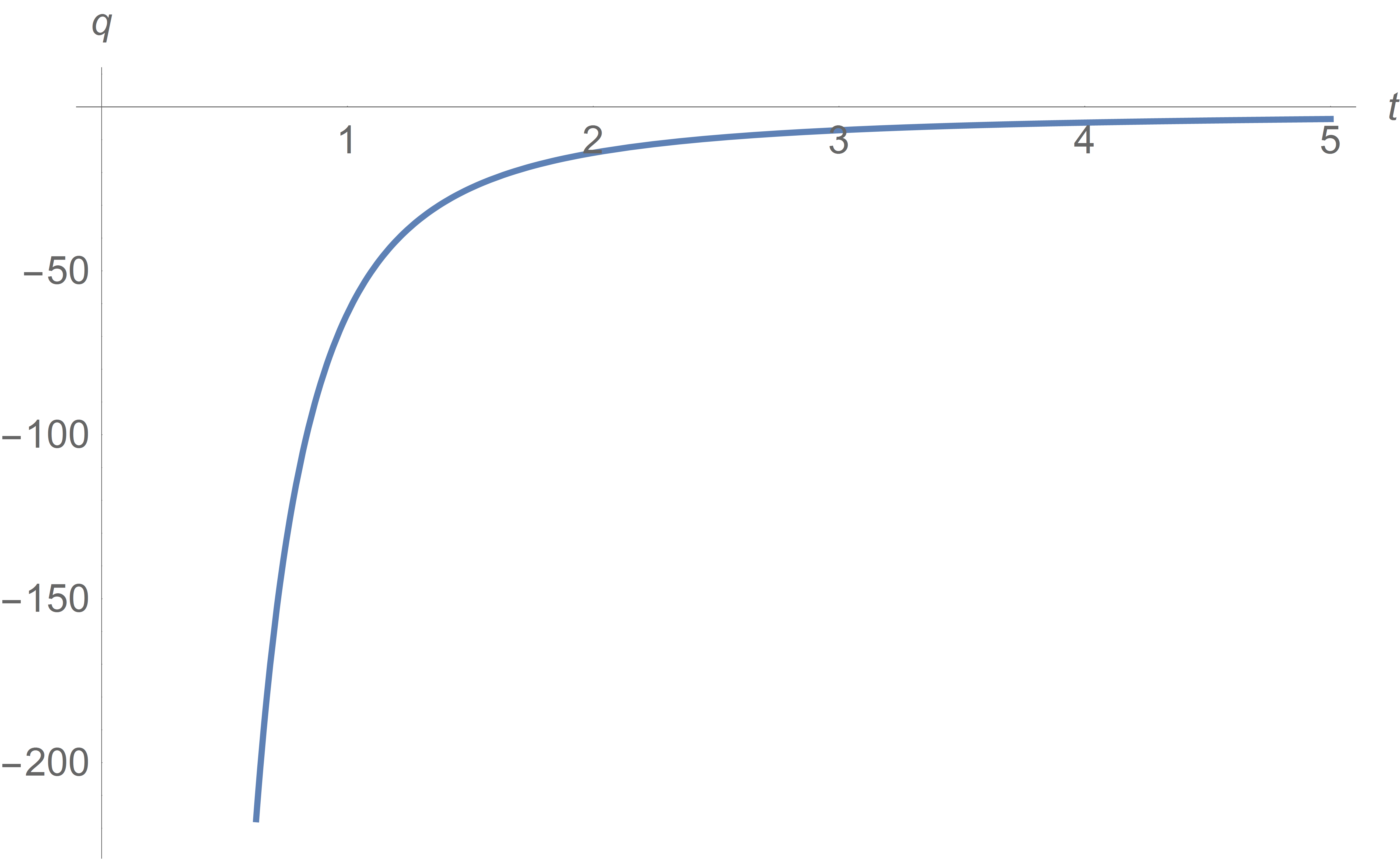}
  \caption{The deceleration parameter $q$ versus cosmic time $t$ in GYrs. The various observational data suggested that \cite{1}, to explain the present cosmic acceleration the range of the deceleration parameter must be lies between -1 to 0. From the figure it is observed that the deceleration parameter $q\to -1$ as $t\to \infty$, which indicates that the expansion of our universe is in an accelerating way. Hence the deceleration parameter is fit with the present observational data \cite{1}}\label{1}
\end{figure}
Expansion factor
\begin{equation}\label{21}
  \theta=3(ct+c_0)
\end{equation}
Recently, Sahni et al. \cite{Sahni} and Alam et al. \cite{Alam} have introduced a pair of new cosmological parameters, the so-called
statefinder parameters $(r, s)$. These are given by
\begin{equation}\label{22}
  r=1+3c(ct+c_0)^{-2}
\end{equation}
and
\begin{equation}\label{23}
  s=-2c[3(ct+c_0)^2+2c]^{-1}
\end{equation}

\begin{figure}
  \centering
  \includegraphics[width=10cm,height=8cm]{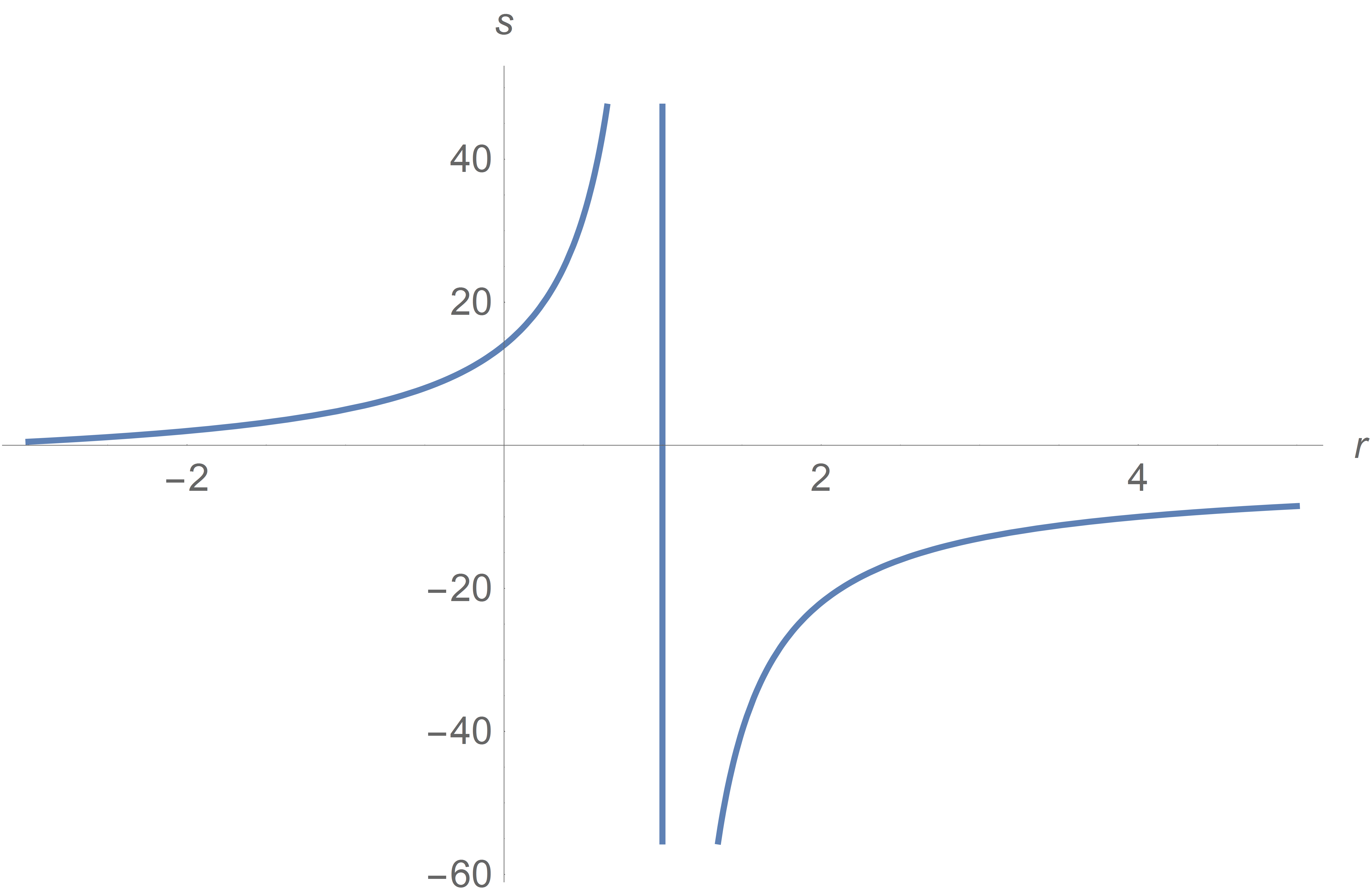}
  \caption{We assumed the cosmic time $t$ in GYrs. From the figure it is observed that the statefinder parameters $r$ and $s$ behave in such a way that $r\to 1$ as $t\to \infty$ and $s\to 0$ as $t\to \infty$, which indicates that our universe represents a $\Lambda CDM$ model in future time \cite{Sahni, Alam}. }\label{2}
\end{figure}
Jerk parameter
\begin{equation}\label{24}
  j(t)=1+\frac{3c}{(ct+c_0)^2}
\end{equation}
And in this case the velocity of sound $c_s$
is given by
\begin{equation}\label{25}
  c_s^2=\frac{dp}{d\rho}=1
\end{equation}
Again from \eqref{2}, \eqref{18} and \eqref{19} we get
\begin{equation}\label{26}
  \rho=a_0e^{3(\omega-1)(\frac{ct^2}{2}+c_0t+c_1)}
\end{equation}
And \eqref{3}, \eqref{18} and \eqref{19} give
\begin{equation}\label{27}
  p=-a_0\frac{\omega+2}{3}e^{3(\omega-1)(\frac{ct^2}{2}+c_0t+c_1)}
\end{equation}
Thus from \eqref{5}, \eqref{26} and \eqref{27} we get
\begin{equation}\label{28}
  \omega=1
\end{equation}
And this implies that the fluid takes the form of the vacuum type of dark energy and the pressure and energy density of the fluid are found to be
constant in this case. However if $\omega\ne 1$, then in this case our universe seems to exist between $t_1$ and $t_2$ given by
\begin{equation}\label{29}
  t=\frac{-2c_0\pm \sqrt{4c_0^2-4c(2c_1-b_1)}}{2c}
\end{equation}
where $b_1=\log\left(\frac{k}{c}\right)$  \\ \\
\textbf{Case-II:}\\
Here we assume the scale factor to be of the form (some call it as the hybrid scale factor)
\begin{equation}\label{30}
  a=t^{\alpha}e^{\beta t},
\end{equation}
where $\alpha$ and $\beta$ are arbitrary constants. Then from \eqref{2} and \eqref{3} we get respectively
\begin{equation}\label{31}
  \rho=\frac{1}{8\pi G}[3(\alpha t^{-1}+\beta)^2+3kt^{-2\alpha}e^{-2\beta t}-\Lambda]
\end{equation}
and
\begin{equation}\label{32}
  p=\frac{1}{8\pi G}[\Lambda-2\{2\alpha\beta t^{-1}+(\alpha^2-\alpha)t^{-2}+\beta^2\}-2(\alpha t^{-1}+\beta)^2-2kt^{-2\alpha}e^{-2\beta t}]
\end{equation}
Now using \eqref{31} and \eqref{11} we get
\begin{equation}\label{33}
  \bigg[3(\alpha t^{-1}+\beta)^2+3k t^{-2\alpha}e^{-2\beta t}-\Lambda\bigg]\frac{\dot{G}}{G}+\dot{\Lambda}=0
\end{equation}
Here taking
\begin{equation}\label{34}
  3(\alpha t^{-1}+\beta)^2+3k t^{-2\alpha}e^{-2\beta t}=b_0\Lambda
\end{equation}
where $b_0$ is an arbitrary constant, equation \eqref{33} gives
\begin{equation}\label{35}
  (b_0-1)\Lambda\frac{\dot{G}}{G}+\dot{\Lambda}=0
\end{equation}
from which we get
\begin{equation}\label{36}
  G=b_1b_0^{b_0-1}[3(\alpha t^{-1}+\beta)^2+3kt^{-2\alpha}e^{-2\beta t}]^{1-b_0}
\end{equation}
Again from \eqref{31} we get
\begin{equation}\label{37}
  \rho=\frac{2}{3\pi}b_1^{-1}b_0^{1-b_0}(1-b_0^{-1})[(\alpha t^{-1}+\beta)^2+kt^{-2\alpha}e^{-2\beta t}] \times [
  (\alpha t^{-1}+\beta)^2+kt^{-2\alpha}e^{-2\beta t}]
\end{equation}
And from \eqref{32} we have
\begin{eqnarray} \label{38}
  p &=& -\frac{1}{4\pi}b_1^{-1}b_0^{1-b_0}[(\alpha t^{-1}+\beta)^2+kt^{-2\alpha}e^{-2\beta t}]^{b_0-1} \\ \nonumber
  &\times&
  \bigg[\left(1-\frac{2}{b_0}\right)\{(\alpha t^{-1}+\beta)^2+kt^{-2\alpha}e^{-2\beta t}\} \\ \nonumber
  &+& \{2\alpha \beta t^{-1}+(\alpha^2-\alpha)t^{-2}+\beta^2\}\bigg]
\end{eqnarray}
In this case we get
\begin{equation}\label{39}
  r=\frac{3\alpha \beta^2 t^2+3(\alpha^2-\alpha)\beta t+ (\alpha-2)(\alpha^2-\alpha)+\beta^3 t^3}{(\alpha+\beta t)^3}
\end{equation}

\begin{figure}
  \centering
  \includegraphics[width=10cm,height=8cm]{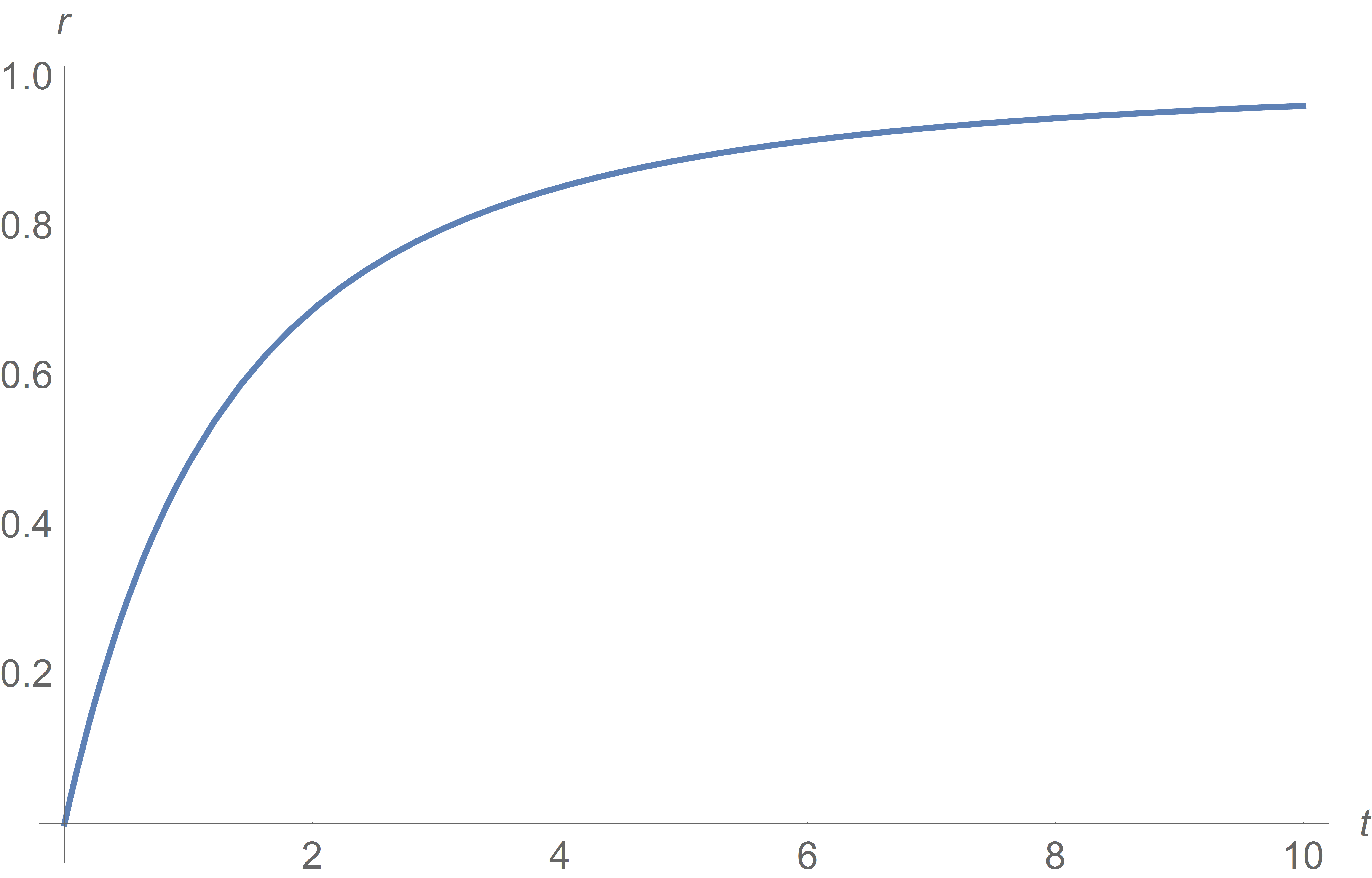}
  \caption{We assumed the cosmic time $t$ in GYrs. From this figure it is observed that the statefinder parameter $r$ varies from zero to one, i. e. $0\le r\le 1$ as $t$ varies from zero to infinity \cite{Sahni, Alam}. }\label{3}
\end{figure}

\begin{equation}\label{40}
  s=\frac{6\alpha \beta t^3 +2\alpha (3\alpha-2)t^2}{9(\alpha+\beta t)^3[2\alpha \beta t+(\alpha^2-\alpha)+\beta^2t^2]}
\end{equation}

\begin{figure}
  \centering
  \includegraphics[width=10cm,height=8cm]{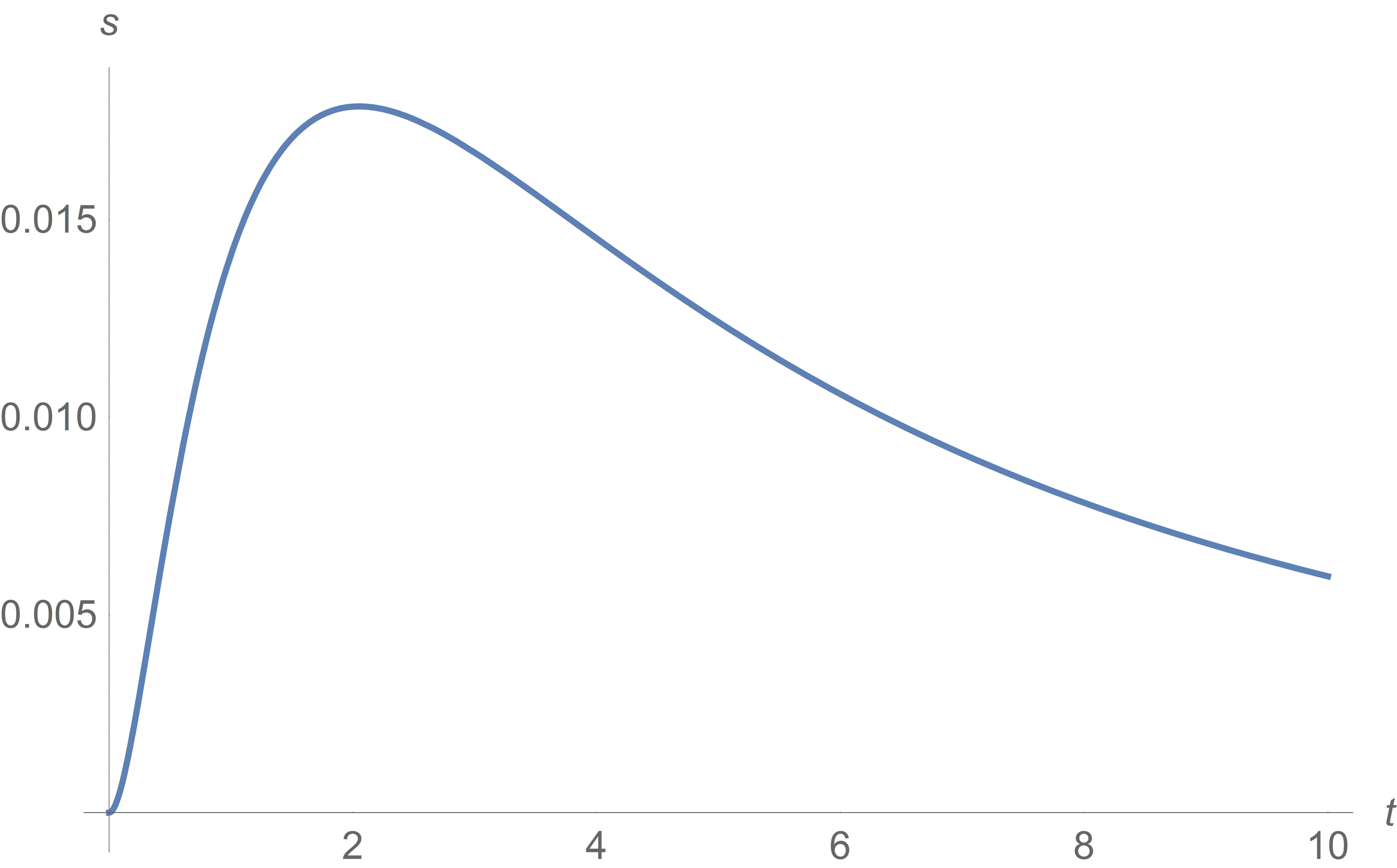}
  \caption{We assumed the cosmic time $t$ in GYrs. From this figure it is observed that the statefinder parameter $s$ varies from zero to some constant which is less than one for finite time; however, after that point $s$ is gradually decreases until it reaches zero as $t\to\infty$ \cite{Sahni, Alam}.}\label{4}
\end{figure}

\begin{equation}\label{41}
  \theta =3(\alpha t^{-1}+\beta)
\end{equation}
\begin{equation}\label{42}
  q=-[2\alpha\beta t^{-1}+(\alpha^2-\alpha)t^{-2}+\beta^2](\alpha t^{-1}+\beta)^2
\end{equation}

\begin{figure}
  \centering
  \includegraphics[width=10cm,height=8cm]{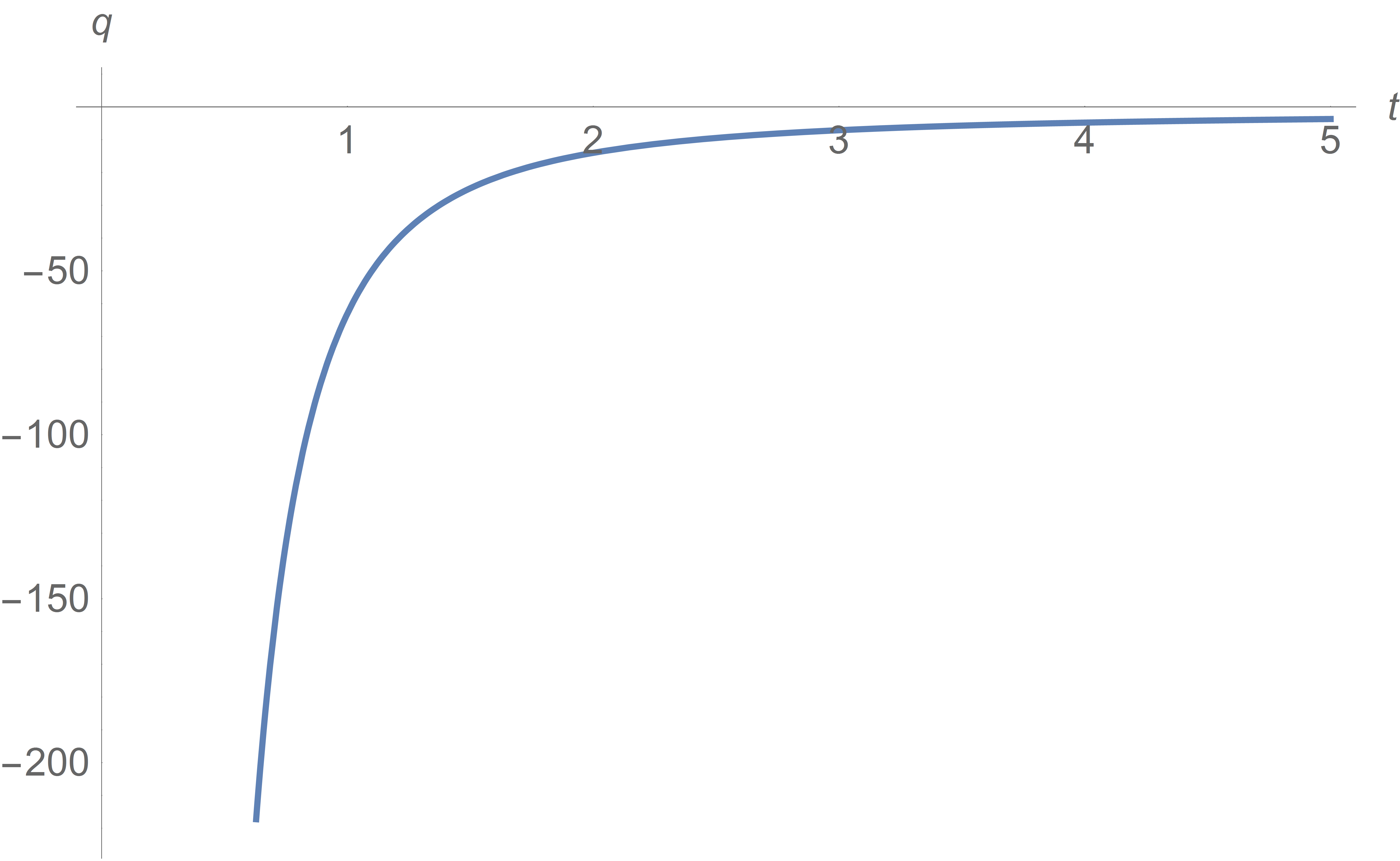}
  \caption{The deceleration parameter $q$ versus cosmic time $t$ in GYrs. The various observational data suggested that \cite{1}, to explain the present cosmic acceleration the range of the deceleration parameter must be lies between -1 to 0. In this case the deceleration parameter $q$ varies from negative infinity to negative one, that is, $-\infty<q<-1$ as $t$ varies from zero to infinity.}\label{5}
\end{figure}

\begin{equation}\label{43}
  j(t)= \frac{3\alpha\beta^2 t^2+3(\alpha^2-\alpha)\beta t+(\alpha-2)(\alpha^2-\alpha)+\beta^3 t^3}{(\alpha+\beta t)^3}
\end{equation}

\textbf{Case-II(a):} \\ \\
As a special case if we take $b_0=2$, then
\begin{equation}\label{44}
  \Lambda=(\alpha t^{-1}+\beta)^2+kt^{-2\alpha}e^{-2\beta t}
\end{equation}

\begin{equation}\label{45}
  G=2b_1[(\alpha t^{-1}+\beta)^2+2kt^{-2\alpha}e^{-2\beta t}]^{-1}
\end{equation}
\begin{equation}\label{46}
  \rho=\frac{3b_1^{-1}}{8\pi}[(\alpha t^{-1}+\beta)^2+kt^{-2\alpha}e^{-2\beta t}]^2
\end{equation}
and
\begin{equation}\label{47}
 p=-\frac{9b_1^{-1}}{8\pi}[(\alpha t^{-1}+\beta)^2+kt^{-2\alpha}e^{-2\beta t}]\times [2\alpha\beta t^{-1}+(\alpha^2-\alpha)t^{-2}+\beta^2]
\end{equation}
Therefore, we have
\begin{equation}\label{48}
  \frac{p}{\rho}=-\frac{(\alpha t^{-1}+\beta)^2-\alpha t^{-2}}{(\alpha t^{-1}+\beta)^2+k t^{-2\alpha}e^{-2\beta t}}
\end{equation}
\\ \\
\begin{figure}
  \centering
  \includegraphics[width=10cm,height=8cm]{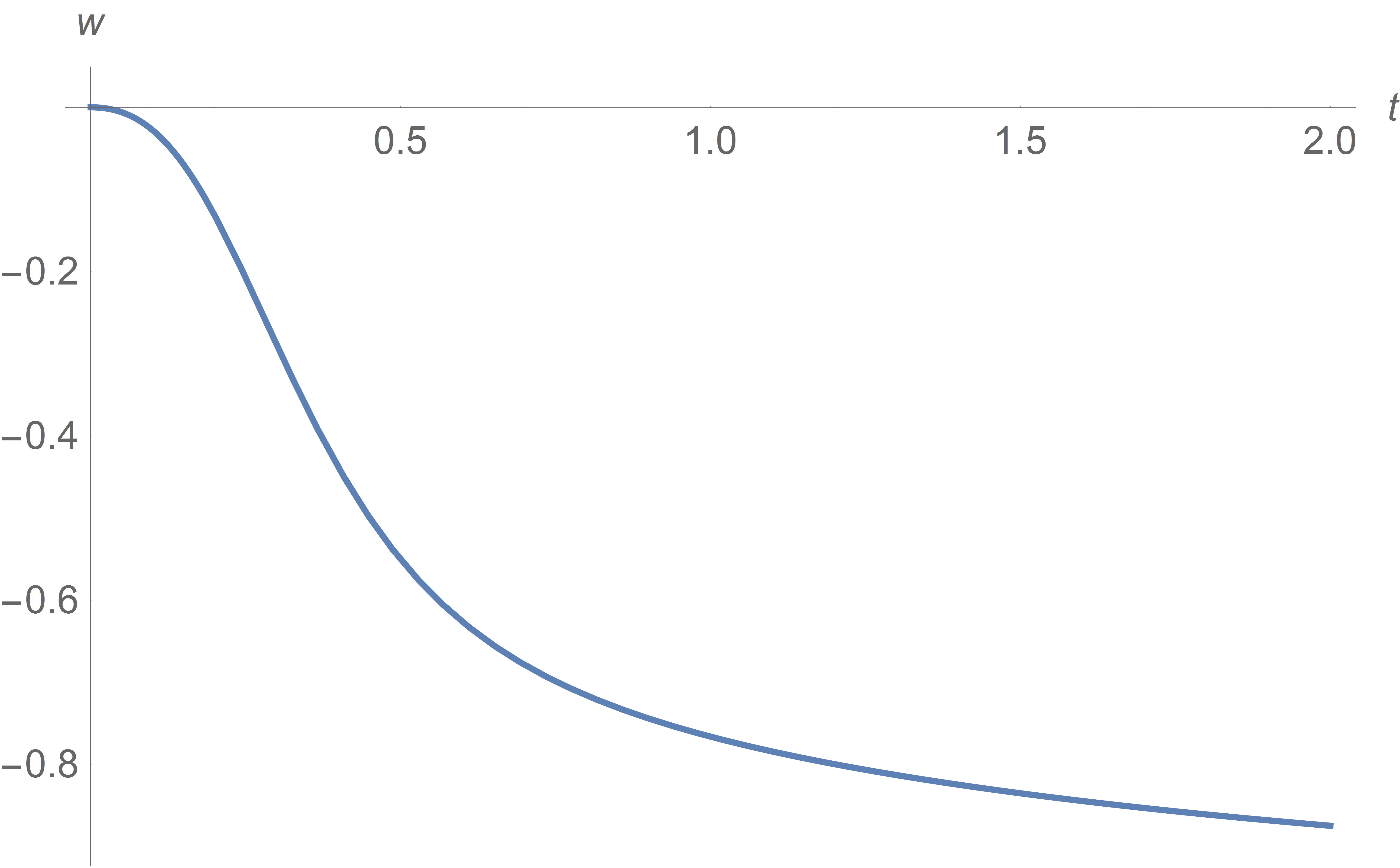}
  \caption{We assumed the cosmic time $t$ in GYrs. In this case the equation of state parameter $\omega$ varies from $-1$ to zero i. e. $-1\le \omega\le 0$. Finally $\omega\to -1$ as $t\to\infty$, which indicates that our universe is dominated by cosmological constant $\Lambda$ \cite{1}.}\label{6}
\end{figure}
\textbf{Case-III:}\\

In this section, we consider the possibility of using a van der Waals fluid equation of state
to describe the present accelerated expansion of the universe. The understanding of the nature of the liquid state of matter
in the universe is connected with the van der Waals equation. This was formulated nearly about 150 years ago \cite{Waals}.
Capozziello et al \cite{Capozziello1}  proposed a model of the universe that contains a baryonic component
with a barotropic fluid equation of state $(p=\omega\rho)$ and a dark fluid component characterized
by a van der Waals equation of state that causes a late time accelerated expansion of the universe.
Unlike conventional barotropic models, a van der Waals fluid contains phase transitions between
different cosmological eras. One fascinating aspect of the van der Waals fluid model
proposed by Capozziello et al \cite{Capozziello1} is that it also contains an early time de-Sitter expansion followed by a
matter dominated epoch without the introduction of a separate scalar field. In other words, it is found that both an early and late time accelerated expansion (e.g. inflation and dark energy) may be caused by the same van der Waals fluid.
Therefore, in this section the model for a spatially flat, homogeneous and isotropic universe is considered, where the source of gravitational field is described by a perfect fluid with a van der Waals equation of state in the absence of dissipative processes \cite{Kremer, Samanta5}. The pressure of the van der Waals fluid $p$ is related to its
energy density $\rho$ by

\begin{equation}\label{49}
  p=\frac{8\omega \rho}{4-\rho}-3\rho^2
\end{equation}
Now from \eqref{4} and \eqref{49} we get
\begin{eqnarray} \label{50}
  \int \frac{da}{a} &=& \frac{1}{12} \int \frac{d\rho}{(\rho-\frac{13}{6})^2+\frac{8\omega}{3}-\frac{121}{36}} \\ \nonumber
   &-& \frac{1}{4(1+2\omega)}\int \frac{d\rho}{\rho}+\frac{1}{8(1+2\omega)}\int \frac{(6\rho-13)d\rho}{
   4+8\omega-13\rho+3\rho^2} \\ \nonumber
   &-& \frac{1}{8(1+2\omega)}\int \frac{13 d\rho}{4+8\omega-13\rho+3\rho^2}
\end{eqnarray}
\textbf{Sub-Case-IIIa:} \\
If $4+8\omega <3 (13/6)^2$, then in this case
\begin{eqnarray}\label{51}
  a &=& b_1 \bigg|\frac{3\rho-\sqrt{121-96\omega}}{3\rho+\sqrt{121-96\omega}}\bigg|^{\frac{12\omega-33}{2(1+2\omega)\sqrt{121-96\omega}}} \\ \nonumber
   &\times& \rho^{-\frac{1}{4(1+2\omega)}}\times (3\rho^2-13\rho+4+8\omega)
\end{eqnarray}
\textbf{Sub-Case-IIIb:} \\
If $4+8\omega =3 (13/6)^2$, then in this case
\begin{equation}\label{52}
  a=b_2 12^{-\frac{6}{169}}\rho^{-\frac{12}{125}}(6\rho-13)^{\frac{12}{169}}e^{\frac{275}{26}(6\rho-13)^{-1}}
\end{equation}
\textbf{Sub-Case-IIIc:} \\
If $4+8\omega >3 (13/6)^2$, then in this case
\begin{eqnarray} \label{53}
  a &=& b_3\rho^{-\frac{1}{4(2\omega+1)}}\times (3\rho^2-13\rho+8\omega+4)^{\frac{1}{8(2\omega+1)}} \\ \nonumber
   &\times& \exp\bigg[{\frac{12\omega-33}{(2\omega+1)(96\omega-121)^{\frac{1}{2}}}\tan ^{-1}\{(96\omega-121)^{\frac{-1}{2}}(6\rho-13)}\}\bigg]
\end{eqnarray}
\textbf{Sub-Case $\omega=0$} \\
It corresponds to the equation of state $p=-3\rho^2$. Thus from \eqref{4} we get
\begin{equation}\label{54}
  a=b_4\rho^{\frac{-1}{4}}(3\rho-1)^{\frac{1}{4}}
\end{equation}

\section{Discussion and Conclusion}
In \textbf{Case-I} we see that the constants $c$ and $c_0$ play important roles in the evolution of the universe, that is, they both determine the
evolutionary behavior as well as ultimate fate of the universe. From the figure-1 it is observed that the deceleration parameter $q\to -1$ as $t\to \infty$, which indicates that the expansion of our universe is in an accelerating way.
Hence the deceleration parameter is fit with the present observational data \cite{1}.
If $c=0$ the statefinder pair becomes $(1, 0)$, which indicates that our universe represents a cosmological model whose equation of state is $\omega=-1$, thus showing that the dark energy filling our universe is of the vacuum energy type \cite{2}. From the figure-2 it is observed that the statefinder parameters $r$ and $s$ behave in such a way that $r\to 1$ as $t\to \infty$ and $s\to 0$ as $t\to \infty$, which indicates that our universe represents a $\Lambda CDM$ model in future time \cite{Sahni, Alam}.

From the expressions of deceleration parameter and expansion factor we see that the rate of expansion of the universe increases with time which is in
agreement with the accelerated expansion of the universe \cite{1}. In this model the pressure and density of the universe happen to be constants. And it is an interesting case when the dark energy dominates other types of energies available in this unverse so that the energy density does not change.

Again on the other hand, if $\omega \ne 1$, then in that case we get the solutions $t_1$ and $t_2$ from \eqref{29} which gives us the cosmological
epoch (between $t_1$ and $t_2$) during which our universe will (exist and) manifest its different activities. And in this case there is less possibility
 of the universe to be a flat one. From the value of the velocity of sound $c_s$ obtained here it is seem that our model universe is a stable one.

In \textbf{Case-II} if $\alpha=0$, then $r=1$ and $s=0$. Thus in this case the statefinder parameter pair $(r, s)$ is found to be $(1, 0)$ which indicates that our universe represents a $\Lambda CDM$ \cite{Sahni, Alam} model with an equation of state parameter $\omega$ equal to $-1$. Also, from the figure-3 it is observed that the statefinder parameter $r$ varies from zero to one, i. e. $0\le r\le 1$ as $t$ varies from zero to infinity. From the figure-4 it is observed that the statefinder parameter $s$ varies from zero to some constant which is less than one for some finite time; however, after that time $s$ is gradually decreases until it reaches zero as $t\to\infty$. Here at $t=0$, the pressure and density of our universe tend to
infinity which indicates that it started its evolution with a big bang. From the expression of $G$ we see that, just after the beginning of the evolution of our universe it inflates until it stops the inflation process for sometime undergoing some different phases until it suddenly bounce to an accelerating
expansion phase, thus $G$ is tending to $0$ at infinitely large time from the beginning.

Again if $\alpha$ is positive and $k\ne -1$, then from \eqref{47} we see that $\frac{p}{\rho}>-1$; thus in this case the dark energy contained in our universe takes the form of quintessence. And in this case $\Lambda$ is not defined at the beginning of the evolution of the universe \cite{Kiefer}, but approaches a finite value $3\beta^2$ as $t \to \infty$. Thus it seems that in this model $\beta$ plays a significant role in its evolutionary history and determine the
ultimate fate of this universe. The various observational data suggested that \cite{1}, to explain the present cosmic acceleration
the range of the deceleration parameter must be lies between -1 to 0. In this case, from the figure-5, it is observed that the deceleration parameter $q$ varies from negative infinity to negative one, that is, $-\infty<q<-1$ as $t$ varies from zero to infinity. Eventually, from the figure-6, it is observed that, the equation of state parameter $\omega$ varies from $-1$ to zero i. e. $-1\le \omega\le 0$. Also, $\omega\to -1$ as $t\to\infty$, which indicates that our universe is dominated by cosmological constant $\Lambda$ \cite{1}.

In \textbf{Case-III} we see that all the three solutions allow values of $\rho$ to exist when $a\to 0$, and asymptotic roots are there whereas the equation
of state is in compromise with the early stage of the universe. Again from equations \eqref{51}, \eqref{52} and \eqref{53} it is found that
\begin{equation}\label{55}
  \rho \to \rho_0=\frac{1}{6}[13-\sqrt{121-96\omega}]
\end{equation}
 which means that, in all the solutions, $\rho_0$ is an asymptotic root. Now from \eqref{55} we obtain
the validity condition in the form $0<\omega \le \frac{21}{96}$ and this is the initial asymptotic condition related to which the universe undergoes expansion as $\rho$ decreases. Thus it is implied that as $a\to 0$, in a fluid with van der Waals equation of state, the energy density does not diverge.
Therefore in this case big bang singularity is not there. On the other hand equations \eqref{49} and \eqref{55} give $p=-\rho_0$. From this we may conclude that for very small value of `$a$' the fluid with van der Waals equation of state contains dark energy of the $\Lambda$-type or the vacuum energy type. And in the beginning of the expansion of our universe the relation $p\approx -\rho$ holds, and thus for $0<\omega\le\frac{121}{96}$, a de Sitter-like expansion is occur. And in the case $\omega=0$ we see that there is a singularity when the energy density of the universe takes the constant value $\frac{1}{3}$, and perhaps it may be a point of bounce. Moreover there is possibility of the universe becoming infinitely flattened as the accelerated expansion increases considerably and thereby the universe ultimately reducing to dust (ending in dust); and this may be considered as a phenomenon due to the dark energy contained in this universe.

Overall, we have studied the late time accelerated expansion of the universe. The exact solution of the Einstein's field equations is derived.
The evolutionary behavior as well as the ultimate fate of the universe, in the course of which the contribution of dark energy in different phases are investigated. At one stage we get a situation (a condition) where the dark energy contained dominates other types of energies available in this universe. In the model universes we obtain here the dark energy is found to be of $\Lambda$CDM and quintessence types-which bear testimony to being real universes. In one of the cases where the equation of state between the fluid pressure and density is of the type of the van der Waals equation, it is found that our universe may end in dust. And, also, it is seen that the behavior of the deceleration parameter is almost compatible with the recent observation.
  The solutions for the scale factor from different models include power law, exponential and product of power law and exponential  universe models. Similar type solutions have been found in the context of a cosmological scenario with
an homogeneous equation of state \cite{Sahni1, Singh1, Singh2, Singh3}. Further, we found that our models do not contain any
finite time singularities in past or future. However, some of the models with time varying cosmological constant $\Lambda$ may allow
finite time singularities in past or future depending on the free parameters of the models discussed by \cite{Supriya}.\\

\textbf{Acknowledgement:} We are very much thankful to the honorable reviewer and Editor for their valuable suggestions for improvement of the presentation of our article. The second author GCS acknowledges (Ref. No. 25(0260)/17/EMR-II) to the CSIR Govt. of India for their support to carry-out the research work.


\begin{thebibliography}{}

\bibitem{1} E. W. Kolb and E. W. Turner, The Early Universe, Addison- Wesley, Redwood City, CA (1990); P. J. E. Peebles, Principle of Physical Cosmology, Princeton Univ. Press, Princeton (1993); J. Peacock, Cosmological Physics, Cambridge University Press, Cambridge, UK (1999);
    Perlmutter et al., Measurements of the cosmological parameters omega and lambda from the first $7$ supernovae at $z>= 0.35$, Ap. J. \textbf{483},
     565 (1997), Discovery of a supernova explosion at half the age of the Universe, Nature \textbf{391}, 51 (1998), Measurements of omega and lambda from $42$ high-redshift supernovae, Ap. J. \textbf{517}, 565 (1999); Schmidt et al., The High-Z supernova search: measuring cosmic deceleration and
     global curvature of the universe using type IA supernovae, Ap. J. \textbf{507}, 46 (1998);
    Riess et al., Observational evidence from supernovae for an accelerating universe and a cosmological constant, Ap. J. \textbf{116}, 1009 (1998).

 \bibitem{2} S. Weinberg, The cosmological constant problem, Rev. Mod. Phys. \textbf{61}, 1 (1989).

 \bibitem{3} A. Einstein, The Principle of Relativity, Methuen (1923), reprinted by
Dover, New York (1924); S. W. Hawking, in General Relativity: An Einstein Centenary Survey,
edited by S. W. Hawking and W. Israel, Cambridge University Press,
Cambridge (1979).

\bibitem{4} P. A. M. Dirac, Proc. R. Soc. A \textbf{165} 119 (1938); \textbf{365} 19 (1979); \textbf{333} 403 (1973); The General
Theory of Relativity (New York : Wiley) 1975.

\bibitem{5} P. S. Wesson (1978), Cosmology and Geophysics (Oxford : Oxford University Press ); P. S.
Wesson (1980), Gravity, Particles and Astrophysics ( Dordrecht : Rieded ).
\bibitem{6} R. K. Tiwari, Astrophys. Space Sci. \textbf{321} 147 (2009); M. Jamil et al, Eur. Phys. J. C \textbf{60} 149
(2009); M. Jamil et al, Phys. Lett. B \textbf{679} 172 (2009); M. R. Setare and M. Jamil, JCAP \textbf{1002}, 010 (2010).

\bibitem{7} F. Hoyle and J. V. Narlikar, Proc. R. Soc. A \textbf{282}, 191 (1964); Nature \textbf{233}, 41 (1971); C. Brans
and R. H. Dicke, Phys. Rev. \textbf{124} 925 (1961).

\bibitem{8} L. M. Krauss and M. S. Turner, The cosmological constant is back, Gen. Rel. Grav. \textbf{27}, 1137 (1995).

\bibitem{Bernstein} J. Bernstein and G. Feinberg, Cosmological Constants (Columbia Univ. Press, New York, 1986), Ch. 1.

\bibitem{Amendola} L. Amendola and S. Tsujikawa, Dark Energy: Theory and Observations, Cambridge University Press, Cambridge UK (2010).

\bibitem{Kiefer} C. Kiefer, F. Queisser and A. A. Starobinsky, Class.Quant.Grav. \textbf{28}, 125022 (2011).

\bibitem{Sahni1} V. Sahni and A. A. Starobinsky, Int.J.Mod.Phys. D \textbf{9},  373 (2000).

\bibitem{Singh1} K. Manihar Singh and K. Priyokumar, Int. J. Theor. Phys. \textbf{53}, 4360 (2014).

\bibitem{Singh2} K. Manihar Singh, K. Priyokumar and M. Dewri, Int. J. Phys. \textbf{3}, 213 (2015).

\bibitem{Singh3} K. Manihar Singh, K. Priyokumar and M. Dewri, Glob. J. Sc. Frontier Res. (A) (Physics and Space Science) \textbf{15}, 15 (2015).

\bibitem{Singh4} K. Manihar Singh et al., Prespacetime Journal, \textbf{7}, 1386 (2016).

\bibitem{Singh5} K. Manihar Singh and K. L. Mahanta, Astrophys. Space Sci. \textbf{361}, 85 (2016).

\bibitem{Priyokumar} K. Priyokumar, K. Manihar Singh and M. R. Mollah, Int. J. Phys. \textbf{56}, 2607 (2017).

\bibitem{Singh6} K. Manihar Singh and Ph. Suranjoy, Int. J. Rec. Tr. Eng. Res. \textbf{3}, 48 (2017).

\bibitem{Supriya} S. Pan, Mod. Phys. Lett. A, \textbf{33}, 1850003 (2018).



\bibitem{Nojiri} S. Nojiri and S. D. Odintsov, eConf C \textbf{0602061}, 06 (2006).

 \bibitem{Nojiri1} S. Nojiri and S. D. Odintsov, Int. J. Geom. Meth. Mod. Phys. \textbf{4}, 115 (2007).

\bibitem{Nojiri2} S. Nojiri and S. D. Odintsov, Phys. Rept. \textbf{505}, 59 (2011).

\bibitem{Felice} A. De Felice and S. Tsujikawa, Living Rev. Rel. \textbf{13}, 3 (2010).

\bibitem{Sotiriou} T. P. Sotiriou and V. Faraoni, Rev. Mod. Phys. \textbf{82}, 451 (2010).

\bibitem{Capozziello} S. Capozziello and M. De Laurentis, Phys. Rept. \textbf{509}, 167 (2011).

\bibitem{Cai} Y. F. Cai, S. Capozziello, M. De Laurentis and E. N. Saridakis, Rept. Prog. Phys. \textbf{79}, 106901 (2016).

\bibitem{Nojiri3} S. Nojiri, S. D. Odintsov and V. K. Oikonomou, Phys. Rept. \textbf{692}, 1 (2017).

\bibitem{Bamba} K. Bamba, Int. J. Geom. Meth. Mod. Phys. \textbf{13}, 1630007 (2016).

\bibitem{Bamba1} K. Bamba and S. D. Odintsov, Symmetry \textbf{7}, 220 (2015).

\bibitem {Bamba2} K. Bamba et al., Mod. Phys. Lett. A \textbf{32}, 1750114 (2017).

\bibitem{Samanta} G. C. Samanta and S. N. Dhal, Int. J. Theor. Phys. \textbf{52}, 1334 (2013)


\bibitem{Samanta1} G. C. Samanta, Int.J.Theor.Phys. \textbf{52}, 2303 (2013).

\bibitem{Samanta2} G. C. Samanta, Int.J.Theor.Phys. \textbf{52}, 2647 (2013).


\bibitem{Samanta3} G. C. Samanta and R. Myrzakulov, Chin.J.Phys. \textbf{55}, 1044 (2017).

\bibitem{Samanta4} G. C. Samanta, R. Myrzakulov and Parth Shah, Z.Naturforsch. A \textbf{72}, 365 (2017).

\bibitem{Oikonomou} V. K. Oikonomou, Mod. Phys. Lett. A, \textbf{32}, 1750172 (2017).

\bibitem{Sahni} V. Sahni, T. D. Saini, A. A. Starobinsky, and U. Alam, JETP Lett. \textbf{77}, 201 (2003).

\bibitem{Alam} U. Alam, V. Sahni, T. D. Saini, and A. A. Starobinsky, Mon. Not. Roy. Astr. Soc. \textbf{344}, 1057 (2003).

\bibitem{Waals} J. D. van der Waals, in Studies in Statistical Mechanics, 14, ed. by J. S. Rowlinson, (North-Holland, Amsterdam, 1988).

\bibitem{Capozziello1} S. Capozziello, S. De Martino, and M. Falanga, Phys. Lett. A, \textbf{299}, 494 (2002).

\bibitem{Kremer} G. M. Kremer, Gen. Rel. Grav. \textbf{36}, 1423 (2004).

\bibitem{Samanta5} G. C. Samanta and R. Myrzakulov, Int. J. Geom. Meth. in Mod. Phys. \textbf{14}, 1750183 (2017).










\end{thebibliography}
\end{document}